\documentclass[a4paper,11pt]{article}

\pdfoutput=1
\usepackage{jcappub}

\usepackage{bbm}
\usepackage{mathrsfs}
\usepackage{slashed}
\usepackage{caption}
\usepackage{epstopdf}
\usepackage[normalem]{ulem}
\usepackage[bottom]{footmisc}
\usepackage{subcaption}
\usepackage{bbold}
\usepackage{titlesec}
\usepackage{threeparttable}
\usepackage{booktabs}
\usepackage{changepage}
\usepackage[utf8]{inputenc}
\usepackage{dsfont} 
\usepackage{grffile}
\usepackage{graphicx}
\usepackage{dcolumn}
\usepackage{bm}
\usepackage{amssymb}
\usepackage{setspace}
\usepackage{amsmath}
\usepackage{array}
\usepackage{float}
\usepackage{lmodern}
\usepackage{multirow}
\usepackage{soul}
\usepackage{braket}
\usepackage{comment}
\usepackage[draft]{pgf}
\usepackage{adjustbox} 
\usepackage{xspace} 
\usepackage{url}
\usepackage{xcolor}
\usepackage{orcidlink}
\usepackage{chngcntr}
\usepackage{indentfirst}

\usepackage{physics}

\usepackage{mathtools,slashed}

\def \p {\partial}

\newcommand{\mi}{\mathrm{i}}
\newcommand{\me}{\mathrm{e}}

\allowdisplaybreaks

\newcommand{\be}{\begin{equation}}
\newcommand{\ee}{\end{equation}}
\def \d {{\rm d}}

\newcommand{\bea}{\begin{eqnarray}}
\newcommand{\eea}{\end{eqnarray}}

\title{Ultralight Boson Ionization from Comparable-Mass Binaries}

\author{
Yuhao Guo\,\orcidlink{0009-0000-8506-8503}$^{a,b,c}$,
Zhen Zhong\,\orcidlink{0000-0002-3138-7530}$^{d,e}$\footnote{Corresponding authors.},
Yifan Chen\,\orcidlink{0000-0002-2507-8272}$^{a,b,c}$\footnotemark[1],
Vitor Cardoso\,\orcidlink{0000-0003-0553-0433}$^{c,e}$,
Taishi Ikeda\,\orcidlink{0000-0002-9076-1027}$^{c}$,
and Lihang Zhou\,\orcidlink{0009-0004-2443-9963}$^{f}$
}

\affiliation[a]{State Key Laboratory of Dark Matter Physics, Tsung-Dao Lee Institute, Shanghai Jiao Tong University, Shanghai 200240, China}
\affiliation[b]{Key Laboratory for Particle Astrophysics and Cosmology (MOE) \& Shanghai Key Laboratory for Particle Physics and Cosmology, Shanghai Jiao Tong University, Shanghai 200240, China}
\affiliation[c]{Center of Gravity, Niels Bohr Institute, Blegdamsvej 17, 2100 Copenhagen, Denmark}
\affiliation[d]{Dipartimento di Fisica, Sapienza Università di Roma \& INFN, Sezione di Roma, Piazzale Aldo Moro 5, 00185, Roma, Italy}
\affiliation[e]{CENTRA, Departamento de F\'{\i}sica, Instituto Superior T\'ecnico -- IST, Universidade de Lisboa -- UL, Avenida Rovisco Pais 1, 1049 Lisboa, Portugal}
\affiliation[f]{Walter Burke Institute for Theoretical Physics, California Institute of Technology, Pasadena, California 91125, USA}

\emailAdd{yhguo21@m.fudan.edu.cn}
\emailAdd{zhen.zhong@uniroma1.it}
\emailAdd{chen.yifan@sjtu.edu.cn}
\emailAdd{vitor.cardoso@nbi.ku.dk}
\emailAdd{m.ikeda.taishi@gmail.com}
\emailAdd{lzhou2@caltech.edu}

\abstract{
Detection of gravitational waves enables probes of environmental effects around compact binaries. Ultralight bosons, well motivated in particle physics and capable of forming core-like dark matter structures, induce environmental dynamics that differ qualitatively from those produced by stars or particle dark matter. For comparable-mass binaries, such bosons can form gravitationally bound states analogous to molecules once the binary separation falls below the characteristic wavelength of the bound states, with an inner region co-moving with the binary. We combine numerical simulations and a semi-analytic framework to characterize the structure and ionization of these gravitational molecules. We determine the extent of the co-moving region and compute the ionization flux driven by orbital motion over a range of eccentricities. Using these results, we estimate the backreaction on the binary orbital evolution and identify a new environmental effect: eccentricity-induced ionization of the co-moving component leads to efficient circularization. We further show that this molecular phase can be astrophysically viable and significantly modify the stochastic gravitational wave background from supermassive black hole binaries.
}

\begin{document}

\maketitle
\flushbottom

\section{Introduction}

The detection of gravitational waves (GWs) has opened a revolutionary window for exploring the universe and compact astrophysical objects, exemplified by observations of stellar-mass black hole (BH) mergers with terrestrial interferometers~\cite{LIGOScientific:2016aoc}  and the detection by pulsar timing arrays (PTAs) of a nanohertz GW background consistent with a population of inspiralling supermassive black hole binaries (SMBHBs)~\cite{NANOGrav:2023gor,EPTA:2023fyk,Reardon:2023gzh,Xu:2023wog,NANOGrav:2023hfp,EPTA:2023xxk}. As we enter the era of precision GW astronomy, an important focus is the study of environmental effects on GW signals, including interactions with stars~\cite{Quinlan:1996vp}, dark matter~\cite{Milosavljevic:2001vi}, and gas~\cite{Gould:1999ia,Armitage:2002uu}. The sensitivity of GW observations, combined with the simplicity of BH systems and the predictive power of general relativity, provides a promising avenue to probe such effects. Recent evidence for a spectral turnover in PTA data suggests the ejection of stars and particle dark matter from SMBHBs, offering a novel way to measure galactic matter densities~\cite{Chen:2024bbg}. A key next question is whether different forms of dark matter produce qualitatively distinct environmental signatures.

Ultralight bosons with sub-eV masses are well-motivated dark matter candidates~\cite{Preskill:1982cy,Abbott:1982af,Dine:1982ah,Svrcek:2006yi,Abel:2008ai,Arvanitaki:2009fg,Goodsell:2009xc}. Owing to their large occupation numbers, they behave as coherently oscillating fields~\cite{Hu:2000ke}. Their interactions with BHs give rise to rich phenomenology, most notably the formation of gravitational atoms, hydrogen-like bound states held together by the BH gravitational potential~\cite{Detweiler:1980uk,Brito:2015oca,Baumann:2019eav}. In BH binaries, however, ultralight boson dynamics become substantially more complex, exhibiting a wide range of behaviors that depend on binary separation and the boson wavelength~\cite{Rozner:2019gba,Blas:2019hxz,Armaleo:2019gil,Ikeda:2020xvt,Annulli:2020lyc,El-Zant:2020god,Liu:2021llm,Broadhurst:2023tus,Aghaie:2023lan,Koo:2023gfm,Bromley:2023yfi,Guo:2023lbv,Aurrekoetxea:2023jwk,Aurrekoetxea:2024cqd,Guo:2024iye,Kim:2024rgf,Tomaselli:2024ojz,Boey:2025qbo,Sarkar:2025tiy,Foster:2025nzf,Chase:2025wwj,Xin:2025ymm,Guo:2025ckp,Roy:2025qaa}.

In this work, we study the dynamics of ultralight bosons around comparable-mass BH binaries, a regime that is fundamentally distinct from analyses focused on gravitational atoms~\cite{Baumann:2018vus,Zhang:2018kib,Zhang:2019eid,Berti:2019wnn,Baumann:2019ztm,Ding:2020bnl,Takahashi:2021eso,Tong:2021whq,DeLuca:2021ite,Su:2021dwz,Takahashi:2021yhy,Baumann:2021fkf,Tong:2022bbl,Baumann:2022pkl,Cole:2022yzw,Kim:2022mdj,Takahashi:2023flk,Tomaselli:2023ysb,Cao:2023fyv,Brito:2023pyl,Fan:2023jjj,Duque:2023seg,Guo:2024iye,Boskovic:2024fga,Tomaselli:2024bdd,Tomaselli:2024dbw,Zhu:2024bqs,Arana:2024kaz,Cao:2024wby,Ginat:2025osk,Peng:2025zca,Kyriazis:2025fis,Li:2025qyu,Tomaselli:2025jfo,Kim:2025wwj,Ding:2025nxe,Tomaselli:2025zdo}. We show that binary-driven ionization of gravitational molecules provides an efficient channel for extracting orbital energy and angular momentum, develop a semi-analytic framework validated by numerical simulations to describe the corresponding ionization fluxes, and demonstrate that this process can induce a characteristic circularization effect and modify the stochastic GW background (SGWB) from supermassive binaries.

The remainder of this paper is organized as follows. In Sec.~\ref{sec:binary-boson}, we present the numerical setup and the basic properties of ultralight boson configurations around comparable-mass BH binaries, including the emergence of gravitationally bound molecular states. Additional details of the numerical implementation are provided in the Appendix (Sec.~\ref{sec:SM}). In Sec.~\ref{sec:ionization}, we develop a semi-analytic framework for the ionization of these molecular states and compare the predicted ionization spectra with numerical simulations. We then turn to astrophysical implications. In Sec.~\ref{sec:formation}, we discuss the formation and survival of gravitational molecules, focusing on dark matter relaxation and related astrophysical viability conditions. In Sec.~\ref{sec:evolution}, we study the backreaction of molecular ionization on the binary orbital evolution and its impact on the SGWB. We conclude in Sec.~\ref{sec:discussion}.

We work with $c=\hbar=1$ and the metric signature $(-,+,+,+)$ throughout this work.

\section{Ultralight Bosons around Comparable-Mass Black Hole Binaries}\label{sec:binary-boson}

We consider a minimally coupled ultralight scalar field $\phi$ interacting only gravitationally in the spacetime of an inspiralling BH binary. Its dynamics are governed by the covariant Klein-Gordon equation,
\be
    \square \phi=\mu^{2}\phi,
\ee
and are simulated with \texttt{GRDzhadzha}~\cite{Aurrekoetxea:2023fhl,Andrade:2021rbd} using the approximate binary BH metric of Ref.~\cite{Bamber:2022pbs},
\be\begin{split}
    \d s^{2}=&-\left(\frac{1+\Phi/2}{1-\Phi/2}\right)^{2}\d t^{2}
    +\left(1-\Phi/2\right)^{4}\left(\d r^{2}+r^{2}\d\theta^{2}+r^{2}\sin^{2}\theta\,\d\varphi^{2}\right), \\
    \Phi=&-\frac{GM}{1+q}\left(\frac{1}{\abs{\vec{r}-\vec{r}_{1}(t)}}+\frac{q}{\abs{\vec{r}-\vec{r}_{2}(t)}}\right),
    \label{eq:metriPhi}
\end{split}\ee
where $(t,r,\theta,\varphi)$ are spherical coordinates centered at the binary's center of mass, with $\theta=\pi/2$ as the orbital plane and the $z$-axis along its normal. Here $\Phi$ is the Newtonian potential sourced by two point masses at $\vec r_1(t)$ and $\vec r_2(t)$, with total mass $M$ and mass ratio $q$, and $G$ is Newton's constant. In this work we adopt an equal-mass binary with $q=1$. We neglect the gravitational potential sourced by the scalar field itself, assuming it is subdominant compared to that of the binary. The metric reduces to Schwarzschild near each BH and to the weak-field limit at large distances.

The binary components follow Keplerian motion. In the center-of-mass frame, assuming the orbital plane lies in the $xy$-plane, their trajectories are parameterized by
\be
\vec r_1(t)=d(t)\frac{1}{1+q}(\cos\beta(t),\sin\beta(t),0),
\qquad
\vec r_2(t)=-d(t)\frac{q}{1+q}(\cos\beta(t),\sin\beta(t),0),
\ee
where $d(t)$ is the binary separation and $\beta(t)$ is the true anomaly. For general eccentric orbits, their evolution is governed by~\cite{Maggiore:2007ulw}
\be
d(t)=a(1-e\cos\mathcal E(t)),
\qquad
\tan^{2}\frac{\beta(t)}{2}=\frac{1+e}{1-e}\tan^{2}\frac{\mathcal E(t)}{2},
\ee
with $a$ and $e$ denoting the semi-major axis and eccentricity, respectively. The eccentric anomaly $\mathcal E(t)$ satisfies
\be
\mathcal E(t)-e\sin\mathcal E(t)=\Omega t,
\ee
where $t=0$ corresponds to pericenter passage.

Analytic expressions for $\beta(t)$ and $d(t)$ can be written as Fourier series~\cite{Maggiore:2007ulw},
\be\begin{split}
\beta(t)&=\Omega t +2\sum_{N=1}^{\infty}\frac{1}{N}\left[\sum_{s=-\infty}^{\infty}J_{N}(-Ne)\Big(\frac{e}{1+\sqrt{1-e^2}}\Big)^{\abs{N+s}}\right]\sin(N\Omega t), \\
d(t)&=a\left(1+\frac{e^2}{2}-2e\sum_{N}\frac{J^{\prime}_{N}(Ne)}{N}\cos(N\Omega t)\right),
\end{split}\ee
where $J_N$ is the Bessel function of the first kind and $J_N'$ its derivative. In the small-eccentricity limit $e\ll 1$, these reduce to
\be
\beta(t)\approx \Omega t+2e\sin\Omega t,
\qquad
d(t)\approx a(1-e\cos\Omega t),
\ee
with Fourier coefficients scaling as $J_N(Ne)\propto e^N$.

As a benchmark, we choose the scalar mass such that the molecular fine-structure constant satisfies $\alpha\equiv\mu\mathcal{M}=0.2$, where $\mathcal{M}\equiv GM$, corresponding to a Bohr radius $r_b=25~\mathcal{M}$. We consider a binary with semi-major axis $a=20~\mathcal{M}$ and define the dimensionless separation $\tilde a\equiv a/r_b$. Molecular bound states are expected to form for $\tilde a\lesssim1$. The resulting orbital frequency and period are $\Omega\simeq0.011~\mathcal{M}^{-1}$ and $T\simeq562~\mathcal{M}$, respectively. The initial scalar field is taken to be a momentarily static spherical Gaussian~\cite{Ikeda:2020xvt} with width $r_b$.

Using \texttt{GRDzhadzha}~\cite{Aurrekoetxea:2023fhl,Andrade:2021rbd}, we employ the standard $3+1$ formalism with sixth-order finite differencing in space and fourth-order Runge–Kutta time integration. The computational domain has a size of $2048~\mathcal{M}$ with ten levels of mesh refinement and a finest grid spacing of $0.0325~\mathcal{M}$. Reflection symmetry is imposed across the equatorial plane, while outgoing radiative boundary conditions are applied at the outer boundaries. Regions inside the BH horizons are excised during the evolution. Further details of the numerical implementation, boundary conditions, and convergence tests are provided in the Supplemental Material.

\begin{figure}[h]
    \centering
     \includegraphics[width=0.9\textwidth]{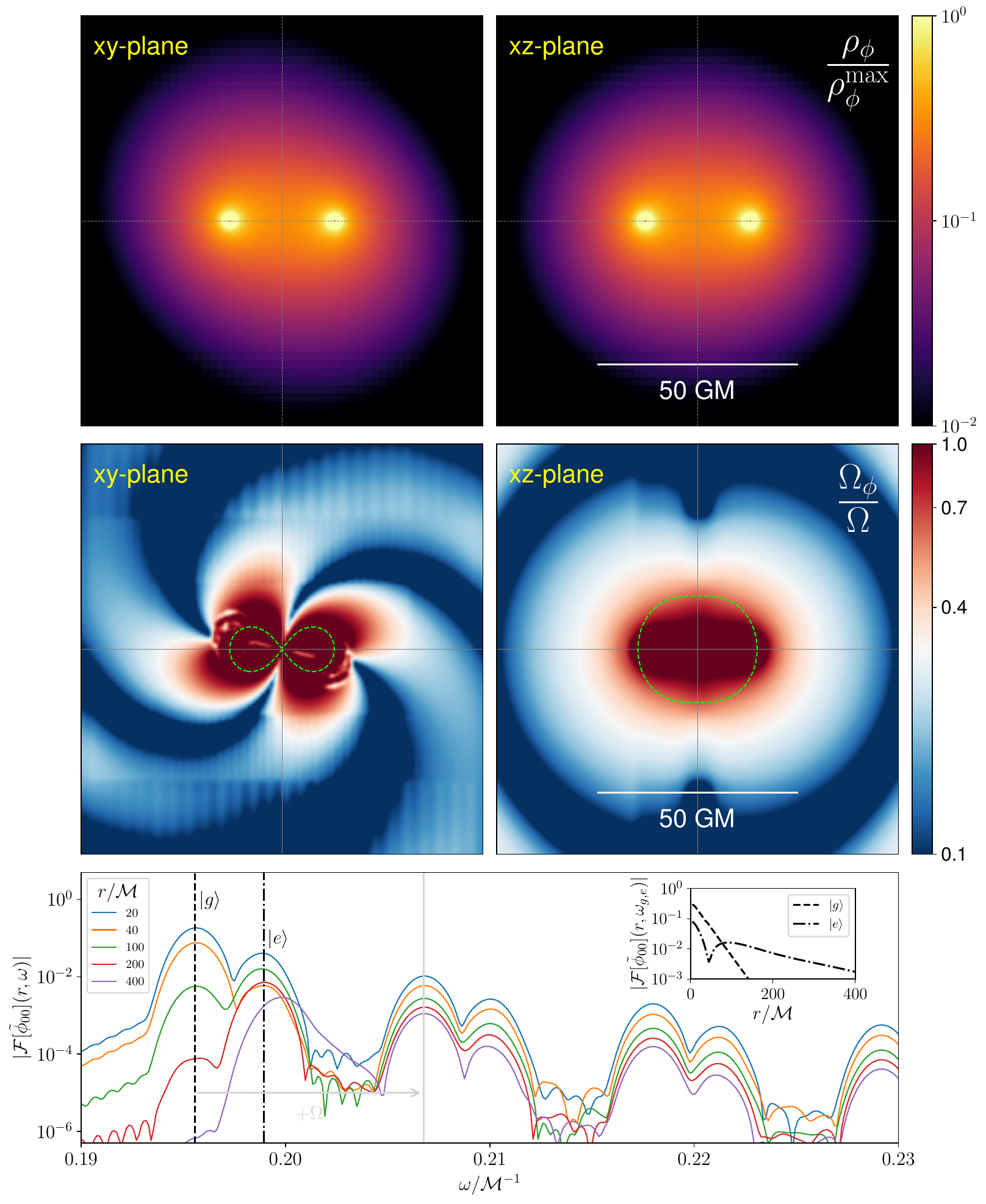}
\caption{
Simulation of a scalar field with mass $\mu \mathcal{M} = 0.2$ around a comparable-mass BH binary with semi-major axis $a = 20\,\mathcal{M}$ and eccentricity $e = 0.3$, shown at the binary's apoapsis.
\textbf{Top:} Distribution of the scalar energy density $\rho$, normalized by its maximum value $\rho_{\rm max}$, on the equatorial $(xy)$ plane and a perpendicular plane, with the $x$-axis aligned along the binary's maximum separation during one orbital period. A movie is available in \cite{movie}.
\textbf{Middle:} Distribution of the scalar angular frequency $\Omega_\phi$, normalized by the orbital frequency $\Omega$, shown at the same time and planes. The green dashed lines indicate the boundary of the co-rotation region, defined by balancing the centrifugal force with the component of the binary's gravitational attraction acting in the opposite direction.
\textbf{Bottom:} Frequency spectrum of the scalar spherical harmonic mode $(\ell, m) = (0, 0)$, with peaks at $\omega\mathcal{M} \approx 0.196$ and $0.199$ corresponding to the ground ($|g\rangle$) and first excited ($|e\rangle$) bound states. The radial wavefunctions, shown in the \textbf{inset panel}, resemble those of gravitational atomic states. Peaks at higher frequencies correspond to ionized states, offset from the bound states by integer multiples of $\Omega$.
}
\label{fig:profile+spectrum}
\end{figure}

We explore several orbital eccentricities. In each scenario, we evolve the scalar field for approximately $10$-$20$ orbits ($\sim10^4\,\mathcal{M}$). Typically, after $3$-$4$ orbital periods ($\sim 2000\,\mathcal{M}$), the scalar system settles into a periodically stable configuration. In Fig.~\ref{fig:profile+spectrum}, we show, for the case with eccentricity $e = 0.3$, snapshots of the scalar energy density $\rho_\phi$ (top), normalized by its maximum value $\rho_\phi^{\rm max}$, and the angular velocity $\Omega_\phi \equiv L_\phi / (\rho_\phi r^2 \sin^2 \theta)$, normalized by the orbital frequency $\Omega$ (middle), both taken at the binary's apoapsis on two perpendicular planes, where $L_\phi$ denotes the angular momentum density of the scalar field. Additionally, we display the frequency spectrum of the spherical harmonic mode $(\ell, m) = (0, 0)$, denoted as $\mathcal{F}[{\Tilde{\phi}}_{00}]$, at various radii (bottom), with $\Tilde{\phi}$ being the dimensionless scalar field value normalized by its initial Gaussian amplitude.

\clearpage

As expected, gravitationally bound states form around the binary. In the frequency spectrum, we observe two peaks at frequencies $\omega\mathcal{M}$ below $\mu\mathcal{M} = 0.2$, approximately matching $\mu (1 - \alpha^2 / 2n^2)$ for $n = 1$ and $2$, respectively. Their radial wavefunctions, shown in the inset panel, confirm that these states closely resemble the ground ($|g\rangle$) and first excited ($|e\rangle$) states of gravitational atoms. At frequencies above $\mu$, multiple peaks appear at frequencies shifted above the bound-state frequencies by integer multiples of $\Omega$, corresponding to ionization waves driven by the binary. These phenomena will be elaborated upon in detail in the subsequent section.

Notably, the inner region of the bound states co-rotates with the binary, having $\Omega_\phi/\Omega \approx 1$ due to the binary drag effect~\cite{Ikeda:2020xvt,Bamber:2022pbs,Aurrekoetxea:2023jwk}. However, at larger distances $r \gg a$, the scalar bound states orbit more slowly than the binary. The boundary of the co-rotating region can be estimated by requiring that the centrifugal force $\propto \Omega^2 r \sin\theta$ balances the binary's gravitational attraction projected in the opposite direction, as illustrated by the green dashed lines. This co-rotation region extends up to a radius of approximately $a/2$, beyond which $\Omega_\phi$ gradually decreases, and exhibits a dipolar structure on the equatorial ($xy$) plane. For eccentric binaries, the boson field exhibits radial oscillations that track the eccentricity-driven radial motion of the binary.

\section{Ionization of Gravitational Molecules by the Binary}\label{sec:ionization}

\begin{figure}[h]
    \centering
    \includegraphics[width=0.7\textwidth]{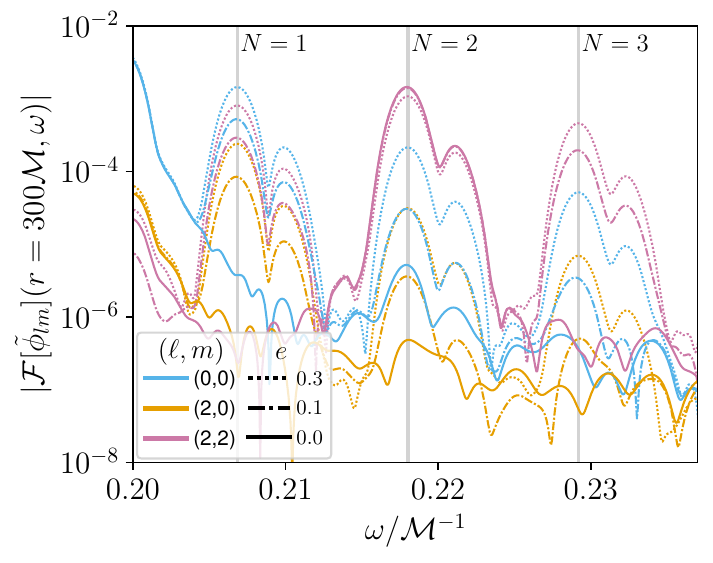}
    \caption{Frequency spectra of the three dominant scalar spherical harmonic modes $(\ell,m)=(0,0)$, $(2,0)$, and $(2,2)$, evaluated at a radius $r_o = 300\,\mathcal{M} \gg r_b$, representing ionization fluxes for various eccentricities $e$. Gray vertical lines indicate ionization frequencies spaced by integer multiples of $N\Omega$ above the ground-state frequency. Circular orbits are dominated by the $(2,2)$ mode at $N=2$, whereas eccentric orbits are dominated by the $(0,0)$ mode at $N=1$, with higher-$N$ modes scaling as $e^N$.}
    \label{fig:spectrum}
\end{figure}

In Fig.~\ref{fig:spectrum}, we show the ionization spectra for different eccentricities $e$, focusing on the three dominant scalar spherical harmonic modes $(\ell,m)=(0,0)$, $(2,0)$, and $(2,2)$ for an equal-mass binary with $q=1$ and $\tilde a=0.8$. The spectra exhibit a series of discrete peaks separated by integer multiples of the orbital frequency $\Omega$, indicating that the ionization process is driven by the periodic motion of the binary.

To explain these features, we now develop a semi-analytic framework for the ionization spectrum. The basic picture is that the bound cloud can be separated into an inner co-moving part, which adiabatically follows the binary motion, and an outer non-co-moving part, which experiences the binary as a time-dependent perturbation. The two regions therefore ionize through different effective potentials and lead to distinct angular and harmonic structures in the emitted spectrum.

We approximate the scalar field $\phi$ as a linear superposition of bound and continuum states. Guided by the frequency spectrum in Fig.~\ref{fig:profile+spectrum} and Fig.~\ref{fig:spectrum}, we retain the dominant ground ($g$) and first excited ($e$) bound states, together with ionized waves of momentum $k$:
\be
\phi(\vec{r},t)=c_g\frac{\psi_{g}(\vec{r}\, )}{\sqrt{2\omega_{g}}}\me^{-\mi\omega_{g}t}
+c_e\frac{\psi_{e}(\vec{r}\,)}{\sqrt{2\omega_{e}}}\me^{-\mi\omega_{e}t}
+\frac{1}{2\pi}\sum_{\ell m}\int \d k \, c_{k\ell m}\frac{\psi_{k\ell m}(\vec{r}\,)}{\sqrt{2\omega_{k}}} \me^{-\mi\omega_{k} t}
+\text{h.c.}+\cdots ,
\ee
where $\cdots$ denotes higher bound states. Here $c$ represents the mode coefficients, $\psi$ the spatial wavefunctions, and $\omega$ the corresponding frequencies.

In the following we focus on ionization from the dominant ground state $|g\rangle$, although the same formalism can be straightforwardly generalized to other initial bound states such as $|e\rangle$. The ionization amplitudes can be computed using Fermi's Golden Rule, in close analogy with gravitational atom ionization~\cite{Baumann:2021fkf,Baumann:2022pkl}:
\be
\begin{split}
    c_{k\ell m}
    =&\, c_{g}\sum_{X=C,\slashed{C}}\sum_{N\in \mathbb{Z}^{+}}
    \eta^{g;X}_{(N)k\ell m}\,
    2\pi\delta(\omega_{k}-\omega_{g}-N\Omega),\\
    \eta^{g;X}_{(N)k\ell m}
    \equiv& \int_{V_X} \psi_{k\ell m}^{\ast}(\vec{r}\,)\,\hat{\mathcal{H}}_{(N)}^{X}(\vec{r}\,)\, \psi_{g}(\vec{r}\,)\,\d^{3}\vec{r},
\label{eq:eta}
\end{split}
\ee
where $X=C,\slashed C$ labels the co-moving and non-co-moving regions, respectively. The quantity $\eta^{g;X}_{(N)k\ell m}$ is the ionization form factor in region $V_X$, and the integer $N$ labels the discrete Fourier components of the effective perturbation,
\be
\hat{\mathcal{H}}^{X}=\sum_{N}\me^{-\mi N \Omega t}\,\hat{\mathcal{H}}_{(N)}^{X}.
\ee
The discrete peaks in the spectrum therefore arise because the periodic binary motion injects energy in integer multiples of the orbital frequency $\Omega$.

A semi-analytic calculation thus requires the initial and final wavefunctions $\psi_g$ and $\psi_{k\ell m}$, the effective perturbing potentials $\hat{\mathcal{H}}_{(N)}^{X}$, and a prescription for separating the co-moving and non-co-moving regions.

\subsection{Initial and Final Wavefunctions}

From Fig.~\ref{fig:profile+spectrum}, the two lowest bound states resemble hydrogenic gravitational atom states in both their frequencies and spatial profiles. We therefore approximate them by isotropic hydrogenic wavefunctions~\cite{Detweiler:1980uk},
\be
\begin{split}
    \psi_{g}(\vec{r}\,)
    &\approx\frac{1}{\sqrt{\pi}r_b^{3/2}}\me^{-\frac{r}{r_b}}, \\
    \psi_{e}(\vec{r}\,)
    &\approx \frac{1}{4\sqrt{2\pi}r_b^{3/2}}\left(2-\frac{r}{r_b}\right)\me^{-\frac{r}{2r_b}},
    \label{eq:bound wf}
\end{split}
\ee
where the molecular Bohr radius is $r_b \equiv 1/(\mu\alpha)$ and the gravitational fine-structure constant is $\alpha \equiv \mu \mathcal{M}$ for boson mass $\mu$. This approximation is justified because the lowest bound states are localized on scales comparable to $r_b$ and closely match the numerically extracted profiles shown in the inset of Fig.~\ref{fig:profile+spectrum} (bottom panel). They satisfy the normalization condition
\be
\int \psi_{g/e}^\ast \psi_{g/e}\, \d^3\vec{r} = 1.
\ee

For the final ionized states, we take non-relativistic hydrogenic continuum spherical waves $\psi_{k\ell m}(\vec{r}\,) \equiv R_{k\ell m}(r)Y_{\ell m}(\theta,\varphi)$ with $\omega_{k\ell m} \approx \mu + k^{2}/(2\mu)$ and radial functions~\cite{Bethe:1957ncq}
\be
\begin{split}
  R_{k\ell m}(r)\approx \frac{1}{r}
  \me^{\frac{\pi \mu \alpha}{2 k}}
  \frac{\abs{\Gamma(\ell+1+\frac{\mi \mu \alpha}{k})}}{(2\ell+1)!}
  (2 k r)^{\ell+1}
  \me^{-\mi k r}
  F_{1} \left(\ell+1+\frac{\mi \mu\alpha}{k}; 2\ell+2; 2\mi kr\right),
  \label{eq: R_klm}
\end{split}
\ee
where $F_{1}$ is the confluent hypergeometric function of the first kind. These states are normalized in momentum space as
\be
\int \psi^{\ast}_{k\ell m}\psi_{k^{\prime}\ell^{\prime}m^{\prime}} \d^3\vec{r}
=2\pi\delta(k-k^{\prime})\delta_{\ell \ell^{\prime}}\delta_{m m^{\prime}}.
\ee
Near the origin, $R_{k\ell m}(r)\to (kr)^{\ell}/r$ as $r\to 0$, while at large distances it asymptotes to a superposition of ingoing and outgoing spherical waves, $R_{k\ell m}(r)\sim(\me^{\mi kr}+\me^{-\mi kr})/r$ as $r\to \infty$. The inclusion of the ingoing component serves as a convenient regularization near the origin~\cite{Baumann:2021fkf,Tomaselli:2023ysb}; the resulting ionization form factors are essentially unchanged compared to the purely outgoing choice. When computing the emitted flux at infinity, only the outgoing component is retained, giving an additional factor of $2$.

As a first approximation, we take the co-moving and non-co-moving regions to be isotropic,
\be
V_C \approx \{ r \leq a \}, \qquad
V_{\slashed{C}} \approx \{ r > a \}.
\ee
Under this simplification, the angular integration in Eq.~(\ref{eq:eta}) can be carried out explicitly, yielding
\be
    \eta^{g;X}_{(N)k\ell m}
    = \int_{V_X} R^{\ast}_{k\ell m}(r)\,
    \left(\hat{\mathcal{H}}_{(N)}^{X} \psi_{g}\right)_{\ell m}
    r^2\d r,
    \qquad X=C,\slashed C,
    \label{eq: reduced eta}
\ee
where $(\cdots)_{\ell m}$ denotes the spherical harmonic component of the enclosed function. A more refined treatment would include the small anisotropic $(\ell,m)=(2,\pm2)$ component of the initial co-moving state and a correspondingly anisotropic co-moving region, which we discuss below.

The spatial wavefunctions $\psi_g$ and $\psi_{k\ell m}$ are the same for both co-moving and non-co-moving ionization. The difference between the two contributions lies entirely in the effective perturbing potentials and the spatial regions over which they act. In the co-moving frame, the coordinate transformation contributes only an additional phase shift $m\Omega t$.

\subsection{External Potentials}

The angular structure of the ionized spectrum can be estimated from the dominant components of $(\hat{\mathcal{H}}_{(N)}^{X}\psi_g)_{\ell m}$. We now discuss the co-moving and non-co-moving contributions in turn.

\subsubsection{Non-co-moving Potential}

We first consider the non-co-moving region, where the cloud experiences the time-dependent Newtonian potential of the orbiting binary,
\be
\hat{\mathcal{H}}^{\slashed{C}} = \mu\Phi,
\ee
with $\Phi$ defined in Eq.~(\ref{eq:metriPhi}). Expanding the Newtonian potential in spherical harmonics~\cite{Martin_1960, Annulli:2020lyc}, one obtains
\be
\begin{split}
    \Phi
    &=\sum_{\ell=0}^{\infty}\sum_{m=-\ell}^{\ell}
    -\frac{G M}{1+q}\me^{-\mi m\beta(t)}
    \frac{4\pi Y_{\ell m}(\frac{\pi}{2},0)}{2\ell+1}
    \left(\mathcal{A}^{1}_{\ell}+q(-1)^{m}\mathcal{A}^{2}_{\ell}\right), \\
    \mathcal{A}^{i}_{\ell}(r,t)
    &\equiv\left(
    \frac{r_{i}^{\ell}(t)}{r^{\ell+1}}\Theta(r-r_{i}(t))
    +\frac{r^{\ell}}{r_{i}^{\ell+1}(t)}\Theta(r_{i}(t)-r)
    \right), \qquad i=1,2.
    \label{eq:Phi}
\end{split}
\ee

For an equal-mass binary ($q=1$), only even nonzero $\ell$ and $m$ contribute to the time-dependent part of the potential for circular orbits. In the extreme mass-ratio limit ($q\ll1$), all $\ell,m\in\mathbb{Z}^{+}$ can contribute, recovering the usual tidal potential discussed in Refs.~\cite{Baumann:2018vus, Baumann:2019ztm}.

Projecting Eq.~(\ref{eq:Phi}) with $q=1$ onto the ground-state wavefunction in Eq.~(\ref{eq:bound wf}), we obtain
\be
\begin{split}
(\hat{\mathcal{H}}_{(N)}^{\slashed{C}}\psi_{g})_{\ell m}
=&-\alpha 
\delta_{m N}
\frac{4\pi Y_{\ell m}(\frac{\pi}{2},0)}{2\ell+1}
\mathcal{A}_{\ell}
\psi_{g},
\qquad \text{for}\ \ell,m\in 2\mathbb{Z}^{+}, \\
\mathcal{A}_{\ell}
\equiv&
\frac{r^{\ell}}{(a/2)^{\ell+1}}\Theta(a/2-r)
+\frac{(a/2)^{\ell}}{r^{\ell+1}}\Theta(r-a/2),
\label{eq: nonco potential}
\end{split}
\ee
at leading order in the eccentricity expansion in the center-of-mass frame. The dominant non-co-moving contribution therefore comes from the $(2,2)$ mode at $N=2$.

\subsubsection{Co-moving Potential}

We next consider the co-moving region. In this region, it is convenient to transform to the co-moving frame, in which the binary positions are fixed and the inner cloud adiabatically follows the orbital motion. For an eccentric orbit, the transformation to co-moving coordinates $(\overline{t}, \overline{r}, \overline{\theta}, \overline{\varphi})$ consists of a radial rescaling together with an azimuthal rotation,
\be
\overline{t}=t,\qquad
\overline{r}=\frac{a}{d(t)}r,\qquad
\overline{\theta}=\theta,\qquad
\overline{\varphi}=\varphi-\beta(t),
\label{eq:coordinate}
\ee
where
\be
d(t) \equiv |\vec{r}_1(t)-\vec{r}_2(t)|\approx a(1-e\cos{\Omega t}),
\qquad
\beta(t)\approx \Omega t + 2e \sin \Omega t
\ee
denote the binary separation and the true anomaly, respectively.

Starting from the Schr\"odinger equation for $\psi_g$, this coordinate transformation induces effective inertial terms in the co-moving frame,
\be
\hat{\mathcal{H}}^{C}
=\left(1-\frac{ \overline{r}^2}{ r^2}\right) \frac{ \overline{\nabla}^2}{2\mu}
+ \mi\frac{\partial \overline{r}}{\partial t} \partial_{\overline{r}}
+ \mi\frac{\partial \overline{\varphi}}{\partial t}\partial_{\overline{\varphi}}
+ \mu{\Phi},
\label{eq:Hc}
\ee
where the first term arises from the transformed kinetic operator, and the next two are inertial terms associated with the time dependence of the new coordinates. The Newtonian potential also acquires a time-dependent component proportional to $(\overline{r}/r-1)$ for eccentric orbits. For a circular orbit, by contrast, $\hat{\mathcal{H}}^{C}$ contains no time-dependent part, so the co-moving contribution vanishes.

Projecting Eq.~(\ref{eq:Hc}) onto $\psi_g$, we find nonvanishing components at $N=1$:
\be
\begin{split}
\label{eq:corotating}
(\hat{\mathcal{H}}_{I(1)}^{C}\psi_{g})_{00}
=&\, e\sqrt{\pi}\left(\frac{\alpha}{\overline{r}}\left(2-\frac{\overline{r}}{r_b}\right) - \Omega \frac{\overline{r}}{r_b}\right)\psi_{g},\\
(\hat{\mathcal{H}}_{\Phi(1)}^{C}\psi_{g})_{\ell m}
=&\, e\alpha \frac{4\pi Y_{\ell m}(\frac{\pi}{2},0)}{2\ell+1}\mathcal{A}_{\ell}\psi_{g},
\qquad \text{for}\ \ell,m\in 2\mathbb{Z}_{\geq 0},
\end{split}
\ee
at leading order in eccentricity and for $q=1$, where $\hat{\mathcal{H}}_{I}^{C}$ and $\hat{\mathcal{H}}_{\Phi}^{C}$ denote the inertial and Newtonian parts, respectively. The dominant co-moving contribution thus comes from the $(0,0)$ mode, with subleading contributions from $(2,0)$ and $(2,2)$. This is qualitatively different from the non-co-moving contribution, which is dominated by $(\ell,m)=(2,2)$.

Physically, this difference reflects the fact that a circular binary generates only quadrupolar time dependence, whereas eccentricity introduces a monopolar breathing mode through the periodic variation of the binary separation. This breathing mode couples directly to the $(0,0)$ channel and explains why the $N=1$ peak in eccentric binaries is dominated by the monopole component. Higher harmonics $N>1$ are further suppressed by higher powers of $e$.

Combining these results with Eq.~(\ref{eq:eta}) and using
\be
|\mathcal{F}[ \tilde{\phi}_{\ell m}]|
\approx \frac{|c_{k \ell m}|\sqrt{2\omega_{k}}}{2\pi kr},
\ee
we estimate the relative amplitudes of the ionization peaks in the frequency spectrum. For a circular orbit, the co-moving contribution vanishes, and the leading signal is the $(2,2)$ mode at $N=2$. For eccentric binaries, the ionization amplitudes scale as $e^N$, making the $N=1$ modes the dominant contribution at small eccentricity. Their predicted relative amplitudes are
\be
|\mathcal{F}[\tilde{\phi}_{00}]|:
|\mathcal{F}[\tilde{\phi}_{20}]|:
|\mathcal{F}[\tilde{\phi}_{22}]|
\approx
6:1:1.
\ee
Higher-order harmonics follow the expected $e^N$ scaling and are therefore increasingly suppressed. The numerical simulation in Fig.~\ref{fig:spectrum} instead gives approximately $6:1:3$. The difference mainly arises from the additional $(2,2)$ component induced by the anisotropy of the co-moving region, as discussed in the next subsection. The overall amplitudes of the $(0,0)$ and $(2,0)$ modes agree well with the numerical spectrum.

\subsection{Anisotropic Structure of the Co-moving Region}

The preceding estimates assume isotropic co-moving and non-co-moving regions separated at $r=a$. This approximation neglects the anisotropy of the co-moving region and therefore underestimates the $(\ell,m)=(2,\pm2)$ component.

A more realistic co-moving region can be estimated classically as the locus where the centrifugal force balances the binary gravitational attraction projected along the centrifugal direction $\hat{n}\equiv(x,y,0)/\sqrt{x^2+y^2}$:
\be
    \mu{\Omega}^{2}\sqrt{x^2+y^2}
    =\frac{\alpha}{2}\left(\frac{\vec{r}_{1}-\vec{r}}{\abs{\vec{r}_{1}-\vec{r}}^{3}\,}+\frac{\vec{r}_{2}-\vec{r}}{\abs{\vec{r}_{2}-\vec{r}\,}^{3}}\right)\cdot\hat{n}.
\ee
We denote this region by $\Sigma_{C}$. As shown by the green contours in Fig.~1, its boundary has a peanut-like shape in the $xy$-plane, implying a non-negligible $(\ell,m)=(2,\pm2)$ component.

Indeed, the angular decomposition of the co-moving region,
$\int_{\Sigma_C} Y_{\ell m}\,\d^3\vec{r}$,
shows that the $(2,2)$ mode contributes at roughly $68\%$ of the $(0,0)$ component. Consequently, the predicted dominant peak ratio becomes $|\mathcal{F}[ \tilde{\phi}_{0 0}]|:|\mathcal{F}[ \tilde{\phi}_{2 0}]|:|\mathcal{F}[ \tilde{\phi}_{22}]|\approx 6:1:2$.
This brings the semi-analytic prediction significantly closer to the numerical result. A fully quantitative comparison with the $N=2$ non-co-moving contribution is not attempted here, since it depends on the precise separation between the co-moving and non-co-moving regions, whose boundary is in reality a smooth transition rather than a sharp interface.

The outermost extent of $\Sigma_C$ reaches only $r\approx a/2$. Our working approximation $V_C\approx\{r\le a\}$ therefore somewhat overestimates the co-moving range. Nevertheless, regions outside $\Sigma_C$ can still have nonzero angular velocity with $\Omega_\phi/\Omega<1$, and should be viewed as mixed regions rather than sharply separated co-moving or non-co-moving domains.

\subsection{Scalings of Ionization Rate}

The ionization rate is related to the form factor by~\cite{Baumann:2021fkf,Baumann:2022pkl}
\be
\Gamma^{X}_{(N)\ell m}=\frac{\mu}{k}\abs{\eta_{(N)\ell m}^{X}}^2,
\qquad X=C,\slashed C.
\label{eq:GammaX}
\ee
We now estimate the scaling of $\Gamma^{C}_{(1)00}$ and $\Gamma^{\slashed{C}}_{(2)22}$ in the limits $\alpha \ll 1$ and $\tilde{a}\equiv a/r_b \ll 1$. The small-$\alpha$ limit justifies the use of hydrogenic wavefunctions in the Newtonian regime and allows us to neglect BH absorption.

For the co-moving contribution, the dominant channel is $(\ell,m)=(0,0)$ at $N=1$. Approximating the perturbation as $\hat{\mathcal{H}}^{C}_{(1)}\sim \alpha/a$ and the initial wavefunction as $\psi_g\sim 1/r_b^{3/2}$, while the continuum radial wavefunction behaves near the origin as $R_{k\ell m}\sim (kr)^{\ell+1}/r$, we have
\be
k \approx \sqrt{2\mu\Omega}
= \sqrt{2}\mu\alpha/\tilde{a}^{3/4},
\qquad
\Omega = \mu\alpha^2/\tilde{a}^{3/2}.
\ee
The radial integration in Eq.~(\ref{eq: reduced eta}) then gives a factor $\sim a^3$, leading to
\be
\Gamma_{(1)00}^{C}\sim \mu\alpha^2 \tilde{a}^{13/4}.
\ee

For the non-co-moving contribution, the dominant channel is $(\ell,m)=(2,2)$ at $N=2$. Taking $\hat{\mathcal{H}}^{\slashed{C}}_{(2)}\sim \alpha/r$, $\psi_g\sim \me^{-r/r_b}/r_b^{3/2}$, $R_{k\ell m}\sim \me^{\mi kr}/r$, and
\be
k \approx \sqrt{2\mu(2\Omega)} = 2\mu\alpha/\tilde{a}^{3/4},
\ee
and evaluating the radial integral in the regime $a \ll r_b \ll 1/k$, we obtain
\be
\Gamma_{(2)22}^{\slashed{C}}\sim \mu\alpha^2 \tilde{a}^{9/4}.
\ee

Thus, although the co-moving contribution controls the $N=1$ eccentric peak through the monopolar breathing mode, the non-co-moving contribution is parametrically less suppressed at small $\tilde a$ and dominates the circular-orbit spectrum through the $(2,2)$ quadrupolar channel.

\begin{figure}[h]
    \centering
     \includegraphics[width = 0.8\textwidth]{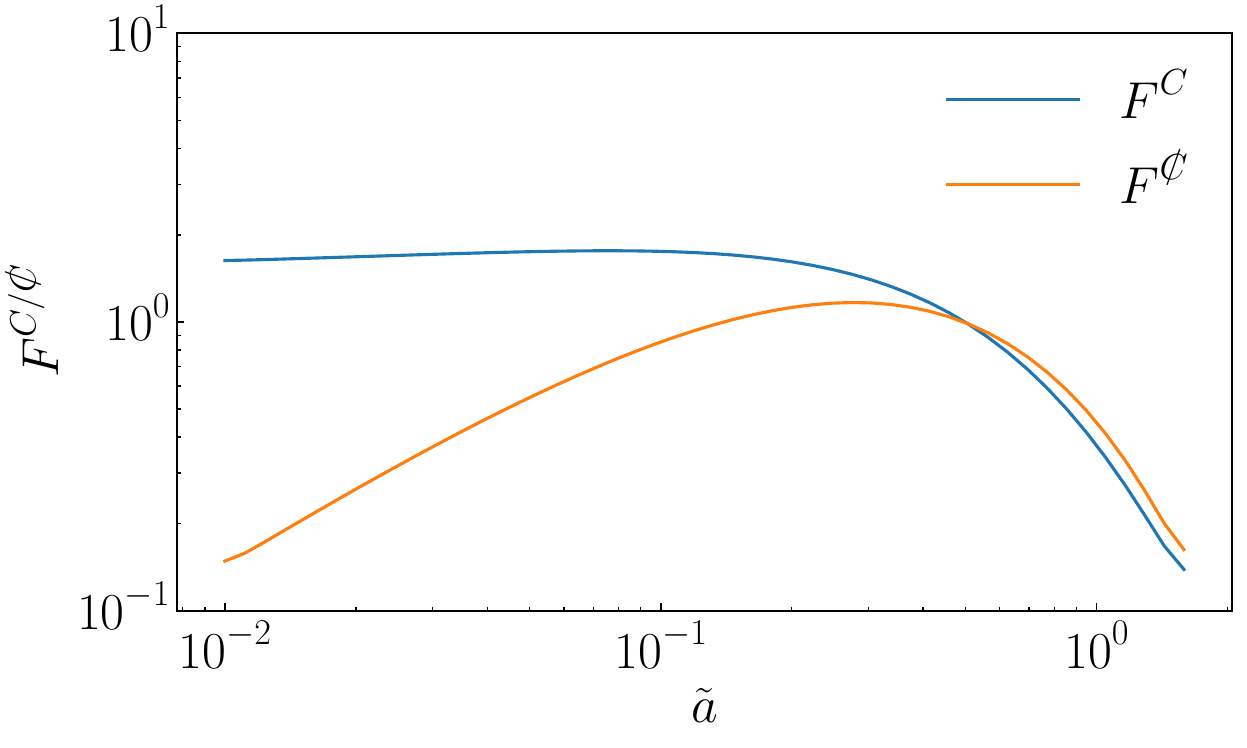}
     \caption{
     Dimensionless coefficients $F^{C}$ and $F^{\slashed{C}}$, defined in Eq.~(\ref{eq:ejection parameter}) and computed from Eq.~(\ref{eq: reduced eta}), as functions of the dimensionless binary separation $\tilde{a}$. They are normalized to unity at $\tilde{a}=0.5$ and exhibit only a mild dependence on $\tilde{a}$. The range $\tilde{a}\in[0,1.6]$ is chosen to include the region where $N=1$ ionization is kinematically allowed.
     }
\label{fig:FC}
\end{figure}

To compare with these analytic estimates, we numerically evaluate Eq.~(\ref{eq: reduced eta}) and parameterize the rates as
\be
\begin{split}
\label{eq:ejection parameter}
\Gamma^{C}_{(1)00}\approx &\, 1.11 \times e^2\mu\alpha^2\tilde{a}^{13/4}F^{C}(\tilde{a}),  \\
\Gamma^{\slashed{C}}_{(2)22}\approx &\, 1.01\times 10^{-3}\mu\alpha^2\tilde{a}^{9/4}F^{\slashed{C}}(\tilde{a}),
\end{split}
\ee
where $F^{C}$ and $F^{\slashed{C}}$ are dimensionless coefficients normalized to unity at $\tilde{a}=0.5$. In the non-relativistic limit considered here, both coefficients are independent of $\alpha$. Their numerical values are shown in Fig.~\ref{fig:FC}.

The weak dependence of $F^{C}$ and $F^{\slashed{C}}$ on $\tilde{a}$ confirms that the leading power-law scalings capture the essential behavior of the ionization rates. Analytic expressions for $F^{\slashed{C}}$ are discussed in Ref.~\cite{Roy:2025qaa}.


\section{Molecular Formation from Dark Matter Accretion}\label{sec:formation}

Given the ionization mechanism and its impact on the molecular cloud, a key question is whether such gravitational molecules can form and be sustained with a sizable mass fraction $M_g/M$ when the binary separation is near the Bohr radius ($\tilde{a} \sim 1$), where $M_g$ denotes the total mass stored in the bound cloud.

 A natural scenario begins with the formation of gravitational atoms from the relaxation of dark matter waves~\cite{Budker:2023sex,Gan:2023swl}, a process particularly efficient around supermassive BHs. For the comparable-mass binaries considered here, both BHs are embedded in the same ultralight-boson environment and therefore are expected to accrete bound clouds at comparable rates. As the binary inspirals, resonant mass transfer between the two gravitational atoms~\cite{Liu:2021llm,Guo:2023lbv,Guo:2024iye,Guo:2025ckp} is expected to redistribute the cloud and establish a molecular configuration before the binary separation reaches the Bohr radius ($\tilde{a}\gtrsim1$). Even if the two BHs initially host clouds of unequal masses, this resonance is expected to occur before the onset of the ionization regime considered in this work.
 Superradiant gravitational atoms~\cite{Penrose:1971uk,ZS,Detweiler:1980uk,Cardoso:2005vk,Dolan:2007mj,Brito:2015oca} provide another possible channel for molecule formation, although it remains unclear whether large clouds can survive tidal disruption and which molecular modes such superradiant states may ultimately occupy.

In the following, we focus on the formation channel through dark matter relaxation and assume that the molecular state has already been established before the binary enters the $\tilde{a}\lesssim1$ regime. We then assess whether the resulting molecular cloud can survive against BH absorption and binary-induced ionization.

Interactions between two free boson waves can reduce the energy of one wave, allowing it to relax into a bound state with negative binding energy. Owing to Bose enhancement, the resulting gravitational atom can grow exponentially, and the ground mode, characterized by a nearly spherical wavefunction, typically dominates~\cite{Budker:2023sex}. Quartic self-interactions, such as those arising from the axion's periodic potential, can facilitate this relaxation, though purely gravitational interactions also suffice. The relaxation timescale in the latter case is estimated as~\cite{Levkov:2018kau}
\be
\tau_{\rm gr}\approx2\times10^{5}\,\text{yrs}\,
\left(\frac{\mu}{10^{-21}\,\text{eV}}\right)
\left(\frac{v_{\rm DM}/c}{0.001}\right)^{6}
\left(\frac{10^{5}\,\text{GeV}/\text{cm}^3}{\rho_{\rm DM}}\right)^{2},
\label{eq:taugr}
\ee
where $\rho_{\rm DM}$ and $v_{\rm DM}$ denote the energy density and typical velocity of the background ultralight boson waves. We focus on the regime in which the de Broglie wavelength of the background waves, scaling as $1/v_{\rm DM}$, is much larger than the Bohr radius, ensuring efficient condensation into bound states.

As shown in Eq.~(\ref{eq:taugr}), the relaxation timescale depends sensitively on the ambient dark matter density. Following Ref.~\cite{Schive:2014dra,Davies:2019wgi}, the central density of a soliton core approximately scales as
\be
\rho_{\rm DM}\approx 10^{5}~\text{GeV}/\text{cm}^{3}
\left(\frac{\mu}{10^{-21}~\text{eV}}\right)^{2}
\left(\frac{v_{\rm DM}/c}{0.001}\right)^{4}.
\ee
As a benchmark, we adopt $M=10^{9}~M_{\odot}$ and $\mu=10^{-21}~\text{eV}$, corresponding to $\alpha=0.01$, together with $v_{\rm DM}/c=0.001$.

To sustain a molecular cloud, gravitational relaxation must compete with both BH absorption and binary-induced ionization. The absorption timescale for the ground mode is~\cite{Detweiler:1980uk}
\be
\tau_{\mathrm{abs}}\approx6.3\times10^{8}\,\text{yrs}\,\left(\frac{M}{10^{9}\,M_{\odot}}\right)\,\left(\frac{0.01}{\alpha}\right)^{6},
\ee
which, for our benchmark parameters, is much longer than the gravitational relaxation timescale.

For ionization, we consider the dominant channel from circular binaries via the $(\ell,m) = (2,2)$ mode at $N=2$ in Eq.~(\ref{eq:ejection parameter}), giving
\be
\tau_{\mathrm{ion}}\approx\, 4.6\times 10^{5}\,\text{yrs}\,\left(\frac{10^{-21}\,\text{eV}}{\mu}\right)\,\left(\frac{0.01}{\alpha}\right)^{2}\,\left(\frac{1}{\tilde{a}}\right)^{9/4}.
\ee
Thus, at large separations ($\tilde a \gtrsim 1$), gravitational relaxation dominates over ionization, allowing the cloud to build up, while ionization becomes competitive only as the binary approaches the molecular regime $\tilde a \sim 1$. Combined with $\tau_{\rm abs} \gg \tau_{\rm gr}$, this hierarchy implies that bound states can be efficiently populated and maintained before being depleted, with BH absorption remaining subdominant.

The above benchmark corresponds to $\alpha=0.01$. To extrapolate to larger values of $\alpha$, we assume the ambient density follows the soliton-core scaling above, giving $\rho_{\rm DM}\propto\mu^2\propto\alpha^2$ for fixed BH mass and velocity dispersion. Consequently, $\tau_{\rm gr}\propto\alpha^{-3}$ and $\tau_{\rm ion}\propto\alpha^{-3}$, while the BH absorption timescale decreases more rapidly, $\tau_{\rm abs}\propto\alpha^{-6}$. Equating $\tau_{\rm gr}$ and $\tau_{\rm abs}$ yields a critical gravitational fine-structure constant,
\be
\alpha_{\rm crit}\approx0.3
\left(\frac{10^{9}\,M_{\odot}}{M}\right)^{2/3}
\left(\frac{v_{\rm DM}/c}{0.001}\right)^{2/3},
\label{eq:alphac}
\ee
below which gravitational relaxation remains the shortest of the three timescales.

Even if dark matter relaxation becomes subdominant at later stages, one can estimate the maximum cloud mass that can be sustained by comparing the orbital energy loss with the energy carried away by ionized waves. Taking the orbital energy difference between $\tilde{a} = 1.6$ and $1$, $\Delta E_{\rm orb}= (M\alpha^2/8) (1-1/1.6)$, and equating it to the ionized energy, $\Delta E_{\rm ion}\approx(\Delta M_g/\mu) \Omega|_{\tilde{a}=1}=\Delta M_g\alpha^2$, yields $\Delta M_g/M \approx 4.7\%$. This estimate therefore indicates that ionization alone is unlikely to completely disrupt the cloud. Throughout this work, we adopt $M_g/M\lesssim10\%$ as a benchmark upper limit, beyond which the cloud’s self-gravity and nonlinear effects, such as Bosenova collapse or self-interaction-induced emission, are expected to invalidate the test-field approximation. Consequently, a substantial molecular cloud can survive throughout the inspiral and continue to influence the binary dynamics.

\section{Binary Orbital Evolution and Gravitational Wave Emissions}\label{sec:evolution}

Having established that a substantial molecular cloud can be formed and maintained, we now study its backreaction on the binary orbit and the resulting GW signal. We focus on the molecular regime $\tilde a \lesssim 1$, where ionization is kinematically allowed once the orbital frequency,
\be
\Omega=\mu\alpha^2/\tilde a^{3/2},
\ee
exceeds the boson binding energy $\mu\alpha^2/2$, which occurs for $\tilde a \lesssim 1.6$. Ionization then extracts both orbital energy and angular momentum from the binary, thereby modifying the evolution of the semi-major axis and eccentricity relative to the standard GW-driven case.

\subsection{Orbital Evolution from Molecular Ionization}

The binary orbital energy and angular momentum are
\be
E = -\frac{GM^2}{8a} = -\frac{M\alpha^2}{8\tilde a}, 
\qquad
L = \frac{\sqrt{GM^3 a (1 - e^2)}}{4}
= \frac{M\sqrt{\tilde a(1-e^2)}}{4\mu}.
\ee
Ionization of the molecular cloud removes energy and angular momentum at rates
\be
\begin{split}
\label{eq:E-L}
    \frac{\d E}{\d t}\Big{|}_{\mathrm{ion}}
    =&-\sum_{X=C,\slashed{C}}\sum_{N\ell m}
    N\Omega \frac{M_{g}}{\mu}\Gamma^{X}_{(N)\ell m}, \\
    \frac{\d L}{\d t}\Big{|}_{\mathrm{ion}}
    =&-\sum_{X=C,\slashed{C}}\sum_{N\ell m}
    m \frac{M_{g}}{\mu}\Gamma^{X}_{(N)\ell m},
\end{split}
\ee
where $\Gamma^{X}_{(N)\ell m}=|\eta^{X}_{(N)\ell m}|^2\mu/k$, defined in Eq.~(\ref{eq:GammaX}), denotes the ionization rate from the co-moving ($X=C$) or non-co-moving ($X=\slashed C$) region. The factor $M_g/\mu$ corresponds to the total boson occupation number in the cloud. Each ionized boson extracts an energy $N\Omega$ and an angular momentum $m$ from the binary.

The orbital evolution of $a$ and $e$ then follows from
\be
\begin{split}
\label{eq:a-e-general}
    \frac{\d a}{\d t}
    = \frac{\d E}{\d t}\left(\frac{\partial E}{\partial a}\right)^{-1},
    \qquad
    \frac{\d e}{\d t}
    =\frac{e^2-1}{2e}
    \left(
    \frac{\d E}{\d t}\frac{1}{E}
    +2\frac{\d L}{\d t}\frac{1}{L}
    \right).
\end{split}
\ee
Substituting Eq.~(\ref{eq:E-L}) and $\Omega=\mu\alpha^2/\tilde a^{3/2}$ gives the general ionization-induced evolution,
\be
\begin{split}
        \frac{\d a}{\d t}\Big{|}_{\mathrm{ion}}
        =& -\frac{M_g}{M}\frac{8\tilde{a}^{1/2}}{\alpha}\frac{1}{\mu}
        \sum_{X=C,\slashed{C}}\sum_{N\ell m}
        N\Gamma^{X}_{(N)\ell m}, \\
    \frac{\d e}{\d t}\Big{|}_{\mathrm{ion}}
    =&\frac{4(1-e^2)}{e}\frac{M_{g}}{M}\tilde{a}^{-1/2}\frac{1}{\mu}
    \sum_{X=C,\slashed{C}}\sum_{N\ell m}
    \left(
    -N +\frac{m}{\sqrt{1-e^2}}
    \right)\Gamma^{X}_{(N)\ell m}.
\end{split}
\ee

As discussed earlier, the dominant ionization channels are the $(\ell,m)=(0,0)$ mode at $N=1$ for the co-moving part and the $(\ell,m)=(2,2)$ mode at $N=2$ for the non-co-moving part; their corresponding rates are given in Eq.~(\ref{eq:ejection parameter}).

Keeping only these dominant channels, the small-eccentricity limit $e\ll1$ yields
\be
\begin{split}
\label{eq:dadtdedt}
    \frac{\d a}{\d t}\Big{|}_{\mathrm{ion}}
    \approx &-8\frac{M_{g}}{M}\alpha \tilde{a}^{11/4}
    \left(\tilde{a}e^2 F^{C} + 2\times10^{-3}F^{\slashed{C}}\right),\\
    \frac{\d e}{\d t}\Big{|}_{\mathrm{ion}}
    \approx & - 4e\frac{M_{g}}{M}\mu\alpha^2 \tilde{a}^{7/4}
    \left(\tilde{a} F^{C} -  10^{-3}F^{\slashed{C}}\right).
\end{split}
\ee

Thus ionization generically drives orbital hardening. For the eccentricity evolution, we work in the small-eccentricity limit $e \ll 1$. The co-moving contribution tends to circularize the orbit, whereas the non-co-moving contribution increases $e$, but becomes dominant only at very small separations, $\tilde a \lesssim 10^{-3}$.

This behavior can be understood from Eq.~(\ref{eq:a-e-general}). The non-co-moving contribution, dominated by the $(\ell,m)=(2,2)$ mode at $N=2$, enters through the combination $(-N + m/\sqrt{1-e^2})$, which evaluates to $(-2 + 2/\sqrt{1-e^2}) \sim e^2$ at leading order. As a result, both the co-moving and non-co-moving contributions to $\d e/\d t$ scale as $e^2$, with $\Gamma^{C}_{(1)00} \propto e^2$.

One may wonder whether subleading terms in the $e$-expansion of the non-co-moving potential could generate contributions at lower order. However, we find that the next-to-leading terms also cancel. In particular, at $N=1$ the potential satisfies $(\hat{\mathcal{H}}^{\slashed{C}}_{(1)})_{00}=0$ and $(\hat{\mathcal{H}}^{\slashed{C}}_{(1)})_{20}=-(\hat{\mathcal{H}}^{\slashed{C}}_{(1)})_{22}$, which implies $\Gamma^{\slashed{C}}_{(1)00}=0$ and $\Gamma^{\slashed{C}}_{(1)20}=\Gamma^{\slashed{C}}_{(1)22}$. Since the radial wavefunctions are independent of $m$, the contributions from $(2,0)$ and $(2,2)$ cancel in Eq.~(\ref{eq:a-e-general}), so that the leading non-co-moving contribution to $\d e/\d t$ vanishes at this order as well. The transition between damping and growth therefore occurs only at $\tilde{a} \sim 10^{-3}$ in the $\tilde{a} \ll 1$, $\alpha \ll 1$ regime.

To determine the full orbital evolution, we combine ionization with the standard GW backreaction~\cite{Peters:1963ux},
\be
\begin{split}
    \frac{\d a}{\d t}\Big{|}_{\mathrm{GW}}
    &=-\frac{16}{5}\frac{G^3M^3}{a^3}
    \frac{1+\frac{73}{24}e^2+\frac{37}{96}e^4}{(1-e^2)^{7/2}}, \\
    \frac{\d e}{\d t}\Big{|}_{\mathrm{GW}}
    &=-\frac{76}{15}\frac{G^3M^3}{a^4}
    \frac{e(1+\frac{121}{304}e^2)}{(1-e^2)^{5/2}}.
\end{split}
\ee
The cloud mass also decreases because of ionization,
\be
\frac{\d M_{g}}{\d t}
= -\sum_{X=C,\slashed{C}}\sum_{N\ell m} \Gamma^{X}_{(N)\ell m} M_g.
\ee
Restricting again to the dominant ionization channels gives
\be
    \frac{\d M_{g}}{dt}\Big{|}_{\mathrm{ion}}
    \approx -M_{g}\mu\alpha^2
    \left(
    \tilde{a}^{13/4}e^2F^{C}(\alpha,\tilde{a})
    + 10^{-3}\tilde{a}^{9/4}F^{\slashed{C}}(\alpha,\tilde{a})
    \right).
\ee
We therefore evolve the coupled system
\be
\begin{split}
    \frac{\d a}{\d t}&= \frac{\d a}{\d t}\Big{|}_{\mathrm{GW}}+ \frac{\d a}{\d t}\Big{|}_{\mathrm{ion}}, \\
    \frac{\d e}{\d t}&= \frac{\d e}{\d t}\Big{|}_{\mathrm{GW}}+ \frac{\d e}{\d t}\Big{|}_{\mathrm{ion}}, \\
    \frac{\d M_{g}}{\d t}&= -\sum_{X=C,\slashed{C}}\sum_{N\ell m} \Gamma^{X}_{(N)\ell m} M_g,
\end{split}
\ee
starting from the onset of the molecular phase at $\tilde a=1$ and varying the initial eccentricity $e_0$ and initial cloud mass $M_g^0$.

Comparing the ionization-induced hardening with the GW-driven inspiral, we find that ionization can dominate at early times for $\tilde a$ slightly below unity. The transition to GW-dominated evolution occurs at a characteristic frequency
\be
\begin{split}
    f_{t}\approx 2.8\,\mathrm{nHz}
    \left(\frac{M_g/M}{0.1}\right)^{0.26}
    \left(\frac{\alpha}{0.05}\right)^{1.7}
    \left(\frac{10^{10}M_{\odot}}{M}\right),
\end{split}
\label{eq:ft}
\ee
where the exponents are decimal approximations of the exact values $6/23$ and $39/23$, obtained from an order-of-magnitude scaling estimate by equating the ionization- and GW-driven orbital evolution in the $e=0$ limit, treating $F^{C}$ as an order-unity constant, neglecting its weak dependence on $\tilde{a}$, and assuming negligible eccentricity at the transition due to the early-stage circularization. The estimate further takes the non-co-moving contribution with $F^{\slashed{C}}\approx1$ to dominate the ionization-induced hardening. The turnover frequency therefore depends primarily on the cloud mass fraction and $\alpha$.


\subsection{Environmental Effects on Supermassive Binaries and the SGWB Turnover}

We now compute the SGWB generated by a simple population of supermassive binaries. For illustration, we adopt the delta-function population
\be
\frac{\d^3\eta}{\d z \d M\d q}
= \delta(M-10^{9}\,M_{\odot})\,\delta(z)\,\delta(q-1)\,{\rm Mpc^{-3}},
\ee
corresponding to equal-mass binaries with $M=10^{9}\,M_{\odot}$ at $z=0$. The SGWB is characterized by the strain spectrum~\cite{Phinney:2001di}
\be
h_{c}^{2}(f)=\frac{4 G}{\pi f}\int \d z \d M \d q \frac{\d^3 \eta}{\d z\d M\d q} \frac{\d E_{\mathrm{GW}}}{\d f_{s}},
\ee
where $\d E_{\mathrm{GW}}/\d f_{s}$ is the GW energy spectrum in the source frame $f_{s}=(1+z)f$. Each binary contributes at orbital harmonics $f_{\mathrm{orb}}^{n}= \Omega/(2\pi)=f_s/n$ for integer $n>0$~\cite{Peters:1963ux},
\be
   \frac{\d E_{\mathrm{GW}}}{\d f_s}
   =\sum_{n=1}^{+\infty}\frac{\d E_{\mathrm{GW}}^{n}/\d t}{n\d f^{n}_{\mathrm{orb}}/dt},
\ee
with
\be
  \frac{\d E_{\mathrm{GW}}^{n}}{\d t}
  =\frac{32G^4 M^5}{5 a^5} \frac{q^2}{(1+q)^4}g(n,e), \qquad
  \frac{\d f^{n}_{\mathrm{orb}}}{\d t}
  =-\frac{3\sqrt{GM}}{4\pi a^{5/2}}\frac{\d a}{\d t},
\ee
where $g(n,e)$ is the standard Peters--Mathews harmonic function,
\be
\begin{split}
    g(n,e)&=\frac{n^4}{32}\Big[\left\{J_{n-2}(ne)-2eJ_{n-1}(ne)+\frac{2}{n}J_{n}(ne)+2eJ_{n+1}(ne)-J_{n+2}(ne)\right\}^2 \\
    &\quad +(1-e^2)\left\{J_{n-2}(ne)-2J_{n}(ne)+J_{n+2}(ne)\right\}^2+\frac{4}{3n^2}J_{n}^2(ne)\Big],
\end{split}
\ee
which reduces to $g(2,0)=1$ for circular orbits. Here $J_n$ denotes the Bessel function of the first kind. For high eccentricities, where a large number of harmonics is required, we adopt the numerical scheme of Ref.~\cite{Chen:2024bbg} to improve efficiency.

\begin{figure}
   \centering
     \includegraphics[width=0.95\textwidth]{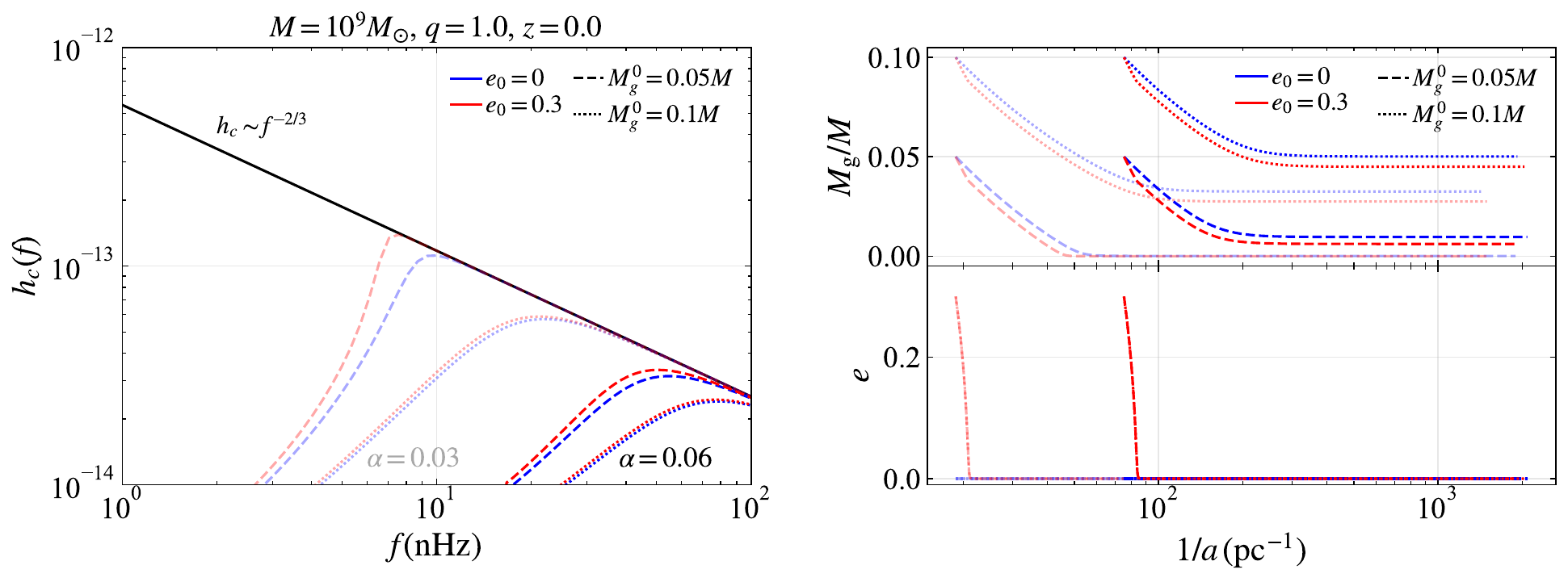}
\caption{\textbf{Left}: SGWB spectra from SMBHB populations for different initial eccentricities $e_0$ and initial bound-state masses $M_g^0$ (both defined at $\tilde{a}=1$), and $\alpha$. The binary population follows a delta-function distribution $\d^3\eta/(\d z \d M\d q) = \delta(M-10^{9}\,M_{\odot})\,\delta(z)\,\delta(q-1)\,{\rm Mpc^{-3}}$. The black line ($h_{c}\sim f^{-2/3}$) corresponds to purely GW-driven circular binaries. Spectra with lower-frequency turnovers (lighter colors) represent $\alpha=0.03$, while those with higher-frequency turnovers correspond to $\alpha=0.06$.
\textbf{Right}: Evolution of the cloud mass fraction $M_g/M$ and orbital eccentricity $e$ as functions of the inverse semi-major axis for the corresponding benchmark models.}
\label{fig:SGWB}
\end{figure}

The resulting SGWB spectra are shown in the left panel of Fig.~\ref{fig:SGWB} for representative values of $\alpha<\alpha_{\rm crit}$ defined in Eq.~(\ref{eq:alphac}), together with different initial eccentricities $e_0$ and cloud masses $M_g^0$, both defined at $\tilde a=1$. The predicted turnover frequencies are in good agreement with the analytic estimate in Eq.~(\ref{eq:ft}). When ionization dominates the binary evolution, the characteristic strain scales approximately linearly with frequency, similar to the case of stellar ejection~\cite{Quinlan:1996vp}. This modified spectral shape can naturally account for a turnover in the nanohertz SGWB spectrum, such as that reported by NANOGrav~\cite{Chen:2024bbg}.

The right panel shows the corresponding evolution of the cloud mass fraction and orbital eccentricity as functions of the inverse semi-major axis. The eccentricity is rapidly damped during the early stage of the evolution, justifying the small-eccentricity approximation adopted in deriving Eq.~(\ref{eq:ft}). During the ionization-dominated phase, the cloud depletion and orbital hardening proceed on comparable timescales, as energy conservation directly relates the orbital energy loss to the energy carried away by the ionized bosons.


\section{Discussion}\label{sec:discussion}

The interplay between ultralight bosons and BH binaries leads to rich phenomenology characterized by distinct physical scales. We have focused on the regime where the binary separation is smaller than the bosonic Bohr radius, resulting in gravitationally bound molecular structures. We have developed a novel calculation of the ionization dynamics for both co-moving and non-co-moving components of gravitational molecules. The resulting backreaction on the binary orbit induces orbital hardening analogous to stellar ejection in three-body systems, yet uniquely accompanied by binary circularization due to the ionization of the co-moving component. The predicted SGWB spectrum from supermassive BH binaries can be directly tested by current PTA observations.

While this study considered purely gravitational interactions, the phenomenology of gravitational molecules would be further enriched by couplings to Standard Model particles. For instance, an axion-photon coupling can lead to observable birefringence signatures~\cite{Chen:2019fsq,Yuan:2020xui,Chen:2021lvo,Chen:2022oad,Ayzenberg:2023hfw}, while quadratic couplings can induce oscillations in the fine structure constant~\cite{Yuan:2022nmu,Bai:2025yxm}. Dense boson clouds can also trigger particle production, such as photons or neutrinos~\cite{Rosa:2017ury,Boskovic:2018lkj,Ikeda:2018nhb,Spieksma:2023vwl,Chen:2023vkq,SHANHE:2024tpr}. Together, these phenomena offer promising targets for multi-messenger astronomy, complementing GW observations with electromagnetic and particle-based probes.

\hspace{5mm}

\begin{acknowledgments}
We are grateful to Katy Clough, Hyungjin Kim, Hidetoshi Omiya, Giovanni Maria Tomaselli, Rodrigo Vicente, Huan Yang, and Hui-Yu Zhu for carefully reading the manuscript and providing valuable comments. We also thank Richard Brito, Joshua Eby, Xucheng Gan, Minyuan Jiang, Yiqiu Ma, Gilad Perez, Daiqin Su, Dina Traykova, Xiao Xue, Jun Zhang, and Rongzi Zhou for insightful discussions.
Z.Z. is supported by the MUR FIS2 Advanced Grant ET-NOW (CUP:~B53C25001080001) and by the INFN TEONGRAV initiative.
The Center of Gravity is a Center of Excellence funded by the Danish National Research Foundation under grant No. 184.
The Tycho supercomputer hosted at the SCIENCE HPC center at the University of Copenhagen was used for supporting this work. This work is supported by VILLUM FONDEN (grant no. 37766), by the Danish Research Foundation, and under the European Union's H2020 ERC Advanced Grant “Black holes: gravitational engines of discovery” grant agreement no. Gravitas-101052587. Views and opinions expressed are however those of the author only and do not necessarily reflect those of the European Union or the European Research Council. Neither the European Union nor the granting authority can be held responsible for them. This project has received funding from the European Union's Horizon 2020 research and innovation programme under the Marie Sklodowska-Curie grant agreement No 101007855. This work was performed in part at Aspen Center for Physics, which is supported by National Science Foundation grant PHY-2210452.
\end{acknowledgments}

\begin{center}
\textbf{\large Appendix: Ultralight Boson Ionization from Comparable-Mass Binaries}
\end{center}
\appendix
\setcounter{equation}{0}
\setcounter{figure}{0}
\setcounter{table}{0}
\numberwithin{equation}{section}
\counterwithin{figure}{section}
\counterwithin{table}{section}

\makeatletter
\@ifpackageloaded{hyperref}{%
  \renewcommand{\theHsection}{app.\Alph{section}}
  \renewcommand{\theHsubsection}{app.\Alph{section}.\arabic{subsection}}
  \renewcommand{\theHsubsubsection}{app.\Alph{section}.\arabic{subsection}.\arabic{subsubsection}}
  \renewcommand{\theHequation}{app.\Alph{section}.\arabic{equation}}
  \renewcommand{\theHfigure}{app.\Alph{section}.\arabic{figure}}
  \renewcommand{\theHtable}{app.\Alph{section}.\arabic{table}}
}{}
\renewcommand{\bibnumfmt}[1]{[#1]}
\renewcommand{\citenumfont}[1]{#1}
\makeatother

\section{Simulation of Boson Field Around a Black Hole Binary} \label{sec:SM}

\subsection{Numerical Implementation}

We solve the Klein-Gordon equation on the binary BH background using the open-source code \texttt{GRDzhadzha}~\cite{Aurrekoetxea:2023fhl,Andrade:2021rbd}, adopting the standard $3+1$ formalism~\cite{Arnowitt:1962hi,Gourgoulhon:2007ue}, where the metric is decomposed as
\be
\d s^2 = g_{\mu\nu} \d x^{\mu} \d x^{\nu} = -\mathcal{N}^2 \d t^2 + \gamma_{ij} (\d x^i + \mathcal{N}^{i} \d t)(\d x^j + \mathcal{N}^{j} \d t),
\ee
with $g_{\mu\nu}$ the spacetime metric, $\mathcal{N}$ the lapse, $\mathcal{N}^{i}$ the shift vector, and $\gamma_{ij}$ the spatial metric.

In this form, the second-order Klein-Gordon equation becomes two coupled first-order equations:
\be
\begin{split}
\partial_{t} \phi &= \mathcal{N} \Pi + \mathcal{N}^{i} \partial_{i} \phi, \\
\partial_{t} \Pi &= \mathcal{N} \gamma^{ij} \partial_{i} \partial_{j} \phi + \mathcal{N} (K \Pi - \gamma^{ij} \mathcal{C}^{k}_{ij} \partial_{k} \phi - \mu^2 \phi) + \partial_{i} \phi \, \partial^{i} \mathcal{N} + \mathcal{N}^{i} \partial_{i} \Pi,
\end{split}
\ee
where $\Pi$ is the conjugate momentum of $\phi$, $K$ is the trace of the extrinsic curvature $K_{ij} = (-\partial_{t} \gamma_{ij} + D_{i} \mathcal{N}_{j} + D_{j} \mathcal{N}_{i})/2\mathcal{N}$, $\mathcal{C}^{k}_{ij}$ are the Christoffel symbols of $\gamma_{ij}$, and $D_i$ is the associated covariant derivative.

For the background metric in Eq.~(\ref{eq:metriPhi}) with $q=1$, we have $\mathcal{N} = (1 + \Phi/2)/(1 - \Phi/2)$, $\mathcal{N}^{i} = 0$, and $\gamma_{ij} = (1 - \Phi/2)^4 \delta_{ij}$.

The initial scalar field configuration is a momentarily static spherical Gaussian:
\begin{eqnarray}
    \phi(\vec{r}\, ,t=0)=Ae^{-\frac{r^{2}}{2\sigma_{0}^{2}}}, \quad \Pi(\vec{r}\, ,t=0)=0,
\end{eqnarray}
with $r \equiv \sqrt{x^2+y^2+z^2}$ the radial coordinate in the center-of-mass frame, and $A$ and $\sigma_{0}$ the field amplitude and Gaussian width, respectively.

To numerically evolve these equations, we employ the method of lines: spatial derivatives are discretized using sixth-order finite-difference stencils, and time integration is carried out with the classical fourth-order Runge-Kutta method. At each timestep we excise the region surrounding each BH center following the procedure of Refs.~\cite{Assumpcao:2018bka,Ikeda:2020xvt}, with details provided in Ref.~\cite{Assumpcao:2018bka}. We impose reflection symmetry across the $z = 0$ plane and apply radiative boundary conditions on all other boundaries, following Refs.~\cite{Alcubierre:2002kk,Clough:2015sqa}:
\be
\begin{split}
\frac{\p \phi}{\p t} = -\left(\sum_{i=1}^{3}\frac{x_i}{r} \frac{\p \phi}{\p x_i}\right) - \frac{\phi}{r}, \\
\frac{\p \Pi}{\p t} = -\left(\sum_{i=1}^{3}\frac{x_i}{r} \frac{\p \Pi}{\p x_i}\right) - \frac{\Pi}{r},
\end{split}
\ee
where $x_i = {x, y, z}$. These conditions correspond to an outgoing relativistic wave of the form $\phi \sim \mathrm{e}^{i(k r - \omega t)}/r$ with $\omega = k$.

For our simulations, we adopt a computational domain of length $L = 2048\,\mathcal{M}$ $(\mathcal{M} \equiv GM)$ with ten levels of mesh refinement. We use fourth-order spatial interpolation to calculate grid variables at finer levels during regridding, and third-order temporal interpolation to obtain intermediate values required for time integration between timesteps. The coarsest grid has spacing $\Delta = 16\,\mathcal{M}$ within a cubic box centered on the binary's center of mass, while the finest grid reaches $\Delta = 0.0325\,\mathcal{M}$. Each external horizon of the binary is resolved with $16$ grid points across its diameter, ensuring adequate accuracy. In addition, the refinement level containing the $r = 300\,\mathcal{M}$ spherical shell used for diagnostic extraction is resolved with spacing $\Delta = 4\,\mathcal{M}$, ensuring accurate measurement.

\subsection{Diagnostic Extraction}

The energy-momentum tensor for a minimally coupled real scalar field is
\be
T_{\mu\nu} = -\frac{1}{2}g_{\mu\nu}\left(\nabla_{\alpha}\phi\nabla^{\alpha}\phi + \mu^2\phi^2\right) + \nabla_{\mu}\phi\nabla_{\nu}\phi.
\ee  
Using the standard $3+1$ decomposition of spacetime, we project this tensor with respect to a normal observer with four-velocity $n^{\mu} = (1/\mathcal{N}, -\mathcal{N}^{i}/\mathcal{N})$:
\be
T_{\mu\nu} = \rho n_\mu n_\nu + S_\mu n_\nu + S_\nu n_\mu + S_{\mu \nu},
\ee  
where  
\be
\rho \equiv n_{\mu}n_{\nu}T^{\mu\nu}, \quad S_{i} \equiv -\gamma_{i\mu}n_{\nu}T^{\mu\nu}, \quad S_{ij} \equiv \gamma_{i\mu}\gamma_{j\nu}T^{\mu\nu}.
\ee  
denote, respectively, the matter energy density, matter momentum density, and matter stress tensor as measured by the normal observer. The spatial metric is $\gamma_{\mu\nu} = g_{\mu\nu} + n_{\mu} n_{\nu}$. Expanding in terms of the scalar variables $\phi$ and $\Pi$ yields~\cite{Clough:2021qlv,Croft:2022gks}
\be
\begin{split}
    \rho &= \frac{1}{2}\left(\Pi^{2} + \gamma^{\mu\nu}\nabla_{\mu}\phi\nabla_{\nu}\phi + \mu^{2}\phi^{2}\right), \\
    S_{i} &= -\Pi\,\gamma_{i}^{\mu}\partial_{\mu}\phi, \\
    S_{ij} &= \frac{1}{2}\gamma_{ij}\left(\Pi^{2} - \gamma^{\mu\nu}\nabla_{\mu}\phi\nabla_{\nu}\phi - \mu^{2}\phi^{2}\right) + \gamma_{i}{}^{\mu}\gamma_{j}{}^{\nu}\nabla_{\mu}\phi\nabla_{\nu}\phi. 
\end{split}
\ee  

The scalar-field energy density $\rho_{\phi}$ and angular-momentum density $L_{\phi}$ are defined as the Noether charges associated with the timelike vector $\zeta_{t}^{\nu} = (1, 0, 0, 0)$ and the rotational vector $\zeta_{\varphi}^{\nu} = (0, -y, x, 0)$ in Cartesian coordinates~\cite{Clough:2021qlv}:
\be
\begin{split}
    \rho_{\phi} &= n_{\nu}\zeta_{t}^{\mu}T_{\mu}^{\nu} = \frac{1}{2}\mathcal{N}\left(\Pi^{2} + \gamma^{\mu\nu}\nabla_{\mu}\phi\nabla_{\nu}\phi + \mu^{2}\phi^{2}\right), \\
    L_{\phi} &= n_{\nu}\zeta_{\varphi}^{\mu}T_{\mu}^{\nu} = -\Pi\,\partial_{\varphi}\phi.
\end{split}
\ee  

We define the angular velocity as
\be
\Omega_{\phi} \equiv \frac{1}{r^{2}\sin^{2}\theta} \frac{\langle L_{\phi} \rangle}{\langle \rho_{\phi} \rangle},
\ee
where $\langle \cdots \rangle$ denotes a time average over the oscillatory behavior of the relevant quantity. In practice, we fit the upper and lower envelopes of the oscillations in $L_{\phi}$ and $\rho_{\phi}$ and take the mean of the two values.

To analyze the field's multipolar content, we decompose $\phi$ into spherical harmonics. We define the dimensionless field $\tilde{\phi} \equiv \phi / A$, normalized by the initial Gaussian amplitude $A$, and extract the spherical-harmonic coefficients at a chosen observation radius $r_o$:
\be
\tilde{\phi}_{\ell m}(r_{o},t) = \iint \tilde{\phi}(r_{o},\theta, \varphi, t) Y^{\ast}_{\ell m}(\theta, \varphi) \,\mathrm{d}\cos\theta \,\mathrm{d}\varphi,
\ee  
where $Y^{\ast}_{\ell m}$ denotes the complex conjugate of the spherical-harmonic function.

\subsection{Convergence Test}

To assess numerical convergence, we examine the time series of $\tilde{\phi}_{00}(r_{o}=20\,\mathcal{M},t)$ for the $e=0.3$ case presented in the main text, focusing on the interval $t \in [0,1000]\,\mathcal{M}$.

\begin{figure}[h]
    \centering
    \includegraphics[width=0.7\linewidth]{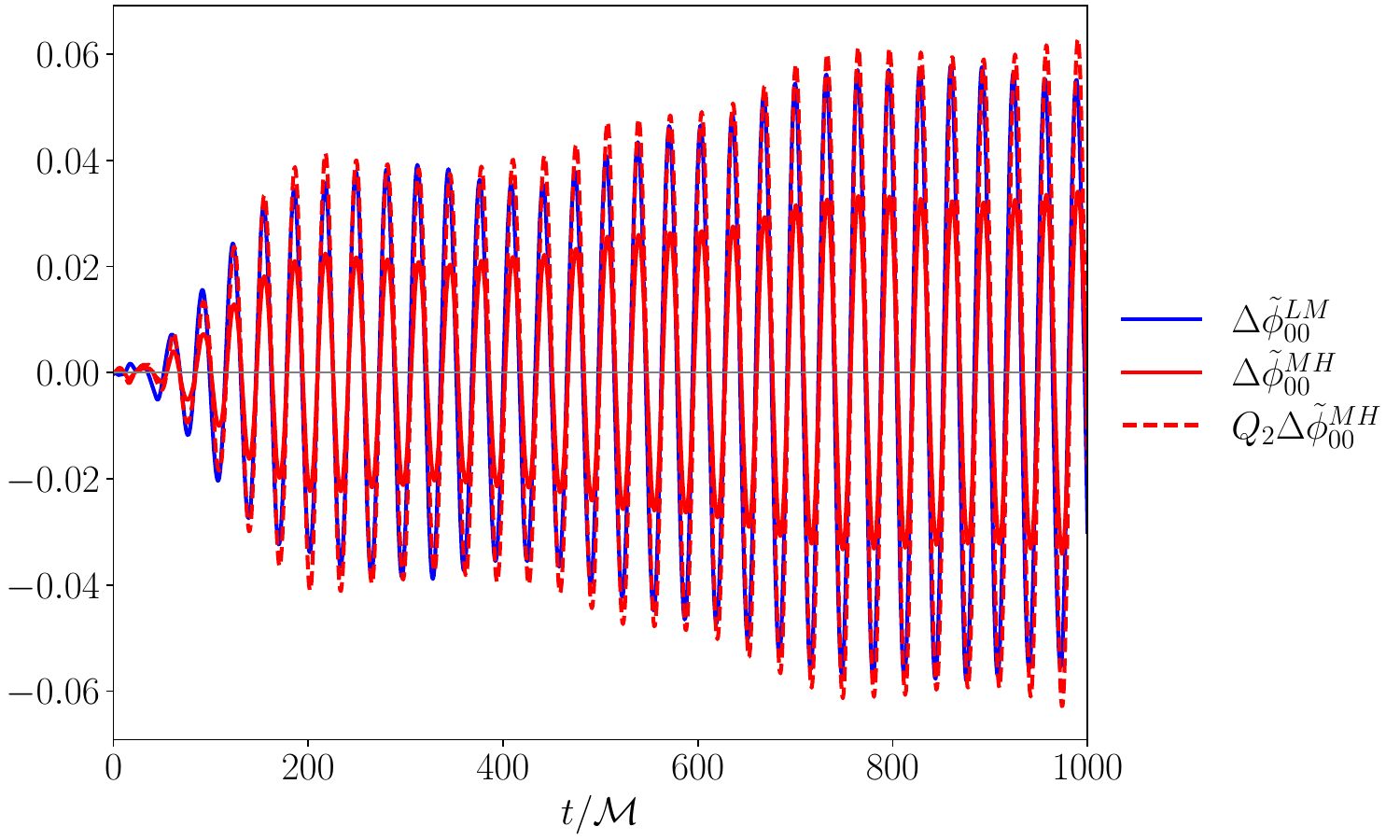}
    \caption{Convergence test of $\tilde{\phi}_{00}(r_{o}=20\,\mathcal{M},t)$ for the $e=0.3$ case during the early stage of the simulation ($t \in [0,1000]\,\mathcal{M}$). Solid lines show the differences between the low–medium resolutions ($\Delta\tilde{\phi}_{00}^{LM}$) and the medium–high resolutions ($\Delta\tilde{\phi}_{00}^{MH}$). The red dashed line shows the rescaled difference $Q_{2}\,\Delta\tilde{\phi}_{00}^{MH}$, indicating approximately second-order convergence.}
    \label{fig:convergence}
\end{figure}

We perform simulations at three resolutions: low, medium, and high, corresponding to coarsest grid spacings of $\Delta_L = 16\,\mathcal{M}$, $\Delta_M = 12.8\,\mathcal{M}$, and $\Delta_H \approx 10.6\,\mathcal{M}$, respectively. In Fig.~\ref{fig:convergence}, we show the differences $\Delta\tilde{\phi}_{00}^{LM}$ (low-medium) and $\Delta\tilde{\phi}_{00}^{MH}$ (medium-high), plotted with solid lines.

To assess the convergence order, we introduce the $o$-th order convergence factor as~\cite{choptuik1999lectures}
\be
Q_{o} \equiv \frac{\Delta_L^o-\Delta_M^o}{\Delta_M^o-\Delta_H^o}.
\ee
We then compare $Q_{o}\,\Delta\tilde{\phi}_{00}^{MH}$ with $\Delta\tilde{\phi}_{00}^{LM}$, using $Q_{2}\approx 1.8$. As shown in Fig.~\ref{fig:convergence}, the rescaled differences follow the expected second-order behavior, indicating convergence within this resolution range.

\providecommand{\href}[2]{#2}\begingroup\raggedright\endgroup


\end{document}